\documentclass[a4paper,twoside]{article}

\usepackage{epsfig}
\usepackage{subfigure}
\usepackage{calc}
\usepackage{amssymb}
\usepackage{amstext}
\usepackage{amsmath}
\usepackage{amsthm}
\usepackage{multicol}
\usepackage{pslatex}
\usepackage{SCITEPRESS}
\usepackage{comment}

\usepackage{bbm}
\usepackage{url}
\usepackage{natbib}
\usepackage{filecontents}

\usepackage{xcolor}
\usepackage{tikz}

\definecolor{green}{HTML}{ABDDA4}
\definecolor{blue}{HTML}{2B83BA}

\begin{document}

\title{Extracting vehicle sensor signals from CAN logs for driver re-identification}

\author{\authorname{Szilvia Lesty{\'a}n\sup{1}, Gergely Acs\sup{1}, Gergely Bicz{\'o}k\sup{1} and Zsolt Szalay\sup{2}}
\affiliation{\sup{1}CrySyS Lab, Dept. of Networked Systems and Services, Budapest Univ. of Technology and Economics, Hungary}
\affiliation{\sup{2}Dept. of Automotive Technology, Budapest Univ. of Technology and Economics, Hungary}
\email{\{lestyan,acs,biczok\}@crysys.hu, zsolt.szalay@gjt.bme.hu}
}


\abstract{
	Data is the new oil for the car industry. Cars generate data about how they are used and  who's behind the wheel which gives rise to a novel way of profiling individuals. 
	Several prior works have successfully demonstrated the feasibility of driver re-identification using the in-vehicle network data captured on the vehicle's CAN (Controller Area Network) bus.
	However, all of them used signals (e.g., velocity, brake pedal or accelerator position)  that have already been extracted from the CAN log  which is itself not a straightforward process. Indeed, car manufacturers intentionally do not reveal the exact signal location within CAN logs.
	Nevertheless, we show that signals can be efficiently extracted from CAN logs using machine learning techniques. We exploit that signals have several distinguishing statistical features which can be learnt and effectively used to identify them across different vehicles, that is, to quasi "reverse-engineer" the CAN protocol. We also demonstrate that the extracted signals can be successfully used to re-identify individuals in a dataset of 33 drivers. Therefore, not revealing signal locations in CAN logs \emph{per se} does not prevent them to be regarded as personal data of drivers.  
}


\onecolumn \maketitle \normalsize \vfill

\section{\uppercase{Introduction}} 
\label{sec:introduction}

Our digital footprint is growing at an unprecedented scale. We use numerous devices and online services creating massive amount of data 24/7. Some of these data are personal, either concerning an identified or and identifiable natural person; thus, they fall under the protection of the European General Data Protection Regulation (GDPR) \cite{eu:gdpr}. In fact, to determine whether a natural person is identifiable based on given data, one should take account of all means reasonably likely to be used (by the data controller or an adversary) to identify the natural person. Such a technique is, e.g., singling out; whether it is reasonably likely to be used depends on the specific data and the scoio-technological context it was collected in \cite{r26gdpr}.

One specific area where digitalization and data generation are booming is automotive. From a set of mechanical and electrical components, cars have evolved into smart cyber-physical systems. Whereas this evolution has enabled automakers to implement advanced safety and entertainment functionalities, it has also opened up novel attack surfaces for malicious hackers and data collection opportunities for OEMs and third parties. The backbone of a smart car is the in-vehicle network which connects ECUs (Electronic Control Units); the most established vehicular network standard is called Control Area Network (CAN) \cite{voss2008comprehensible}. CAN is already a critical technology worldwide making automotive data access a commodity. One or more CAN buses carry all important driving related information inside a car. OEMs (Original Equipment Manufacturers, i.e., car makers) collect and analyze CAN data for maintenance purposes; however, CAN data might reveal other, more personal traits, such as the driving behavior of natural persons. Such information could be invaluable to third party service providers such as insurance companies, fleet management services and other location-based businesses (not to mention malicious entities), hence there exist economic incentives for them to collect or buy them.

It has been shown that automobile driver fingerprinting could be practical based on sensor signals captured on the CAN bus in restricted environments \cite{fingerprint}. Using machine learning techniques, authors re-identified drivers from a fixed set of experiment participants, thus implementing singling out, which makes this a privacy threat. There is a caveat: the adversary has to know the higher layer protocols of CAN in order to extract meaningful sensor readings. Since such message and message flow specifications (above the data link layer) are usually proprietary and closely guarded industrial secrets, such adversarial background knowledge might not be reasonable. In this case, the research question changes: is it possible for an adversary to re-identify drivers based on raw CAN data without the knowledge of protocols above the data link layer?

\noindent \textbf{Contributions. }
In this paper we investigate experimentally the potential to identify and extract vehicle sensor signals from raw CAN bus data for the sake of inferring personal driving behavior and re-identifying drivers. As signal positions, lengths and coding are proprietary and vary among makes, models, model years and even geographical area, first, we have to interpret the messages.
We emphasize that we do not intend to perform (an even remotely) comprehensive reverse engineering \cite{reverse_surv}; we focus solely on a small number of sensor signals which are good descriptors of natural driving behavior.  

Our contributions are three-fold:
\begin{enumerate}
    \item we devise a heuristic method for message decomposition and log pre-processing;
    \item we build, train and validate a machine learning classifier that can efficiently match vehicle sensor signals to a ground truth based on raw CAN data. In particular, we train a classifier on the statistical features of a signal in one car (e.g., Opel Astra), then we use this trained classifier to localize the same signal in a different car (e.g., Toyota). The intuition is that the physical phenomenon represented by the signal has identical statistical features across different cars, and hence can be used to identify the same signal in all cars using the same classifier;
    \item we briefly demonstrate that re-identification of drivers is possible using the extracted signals.
\end{enumerate}

The rest of the paper is structured as follows. Section \ref{sec:related} presents related work. Section \ref{sec:can} gives a background on important characteristics of the Controller Area Network. Section \ref{sec:data} describes our data collection process. Section \ref{sec:analysis} presents our efforts on message decomposition and log pre-processing. Section \ref{sec:classification} presents the design, evaluation and validation of our random forest classifier for extracting sensor signals. Section \ref{sec:reid} briefly demonstrates the successful application of the extracted signals for driver re-identification. Finally, Section \ref{sec:conclusion} concludes the paper.

\section{\uppercase{Related work}}
\label{sec:related}

Driver characterization based on CAN data has gathered significant research interest from both the automotive and the data privacy domain. The common trait in these works is the presumed familiarity with the whole specific CAN protocol stack including the presentation and application layers giving the researchers access to sensor signals. This knowledge is usually gained via access to the OEM's documentations in the framework of some research cooperation. As such, researchers do not normally disclose such information to preserve secrecy.

Miyajima et al. has investigated \cite{GMM_driving} driver characteristics when following another vehicle and pedal operation patterns were modeled using speech recognition methods. Sensor signals were collected in both a driving simulator and a real vehicle. Using car-following patterns and spectral features of pedal operation signals authors achieved an identification rate of 89.6\% for the simulator (12 drivers). For the field test, by only applying cepstral analysis on pedal signals the identification rate was down to 76.8\% (276 drivers). Fugiglando et al.~\cite{drivers_classification} developed a new methodology for near-real-time classification of driver behavior in uncontrolled environments, where 64 people drove 10 cars for a total of over 2000 driving trips without any type of predetermined driving instruction. Despite their advance use of unsupervised machine learning techniques they conclude that clustering drivers based on their behavior remains a challenging problem. 

Hallac et al. \cite{hallac2016driver} discovered that driving maneuvers during turning exhibit personal traits that are promising regarding driver re-identification. Using the same dataset from Audi and its affiliates, Fugiglando et al. \cite{driving_dna}, showed that four behavioral traits, namely braking, turning, speeding and fuel efficiency could characterize driver adequately well. They provided a (mostly theoretical) methodology to reduce the vast CAN dataset along these lines.

Enev et al. authored a seminal paper \cite{fingerprint}  which makes use of mostly statistical features as an input for binary (one-vs-one) classification with regard to driving behavior. Driving the same car in a constrained parking lot setting and a longer but fixed route, authors re-identified their 15 drivers with 100\% accuracy. Authors had access to all available sensor signals and their scaling and offset parameters from  the manufacturer's documentation. 

In a paper targeted at anomaly detection in in-vehicle networks~\cite{field_class}, authors developed a greedy algorithm to split the messages into fields and to classify the fields into categories: constant, multi-value and counter/sensor. Note that the algorithm does not distinguish between counters and sensor signals, and the semantics of the signals are not interpreted. Thus, their results cannot be directly used for inferring driver behavior.

\section{CAN: Controller Area Network}
\label{sec:can}



The Controller Area Network (CAN) is a bus system providing in-vehicle communications for ECUs and other devices. The first CAN bus protocol was developed in 1986, and it was adopted as an international standard   in 1993 (ISO 11898). A recent car can have anywhere from 5 up to 100 ECUs, which are served by several CANs. Our point of focus is the CAN serving the drive-train.

CAN is an overloaded term \cite{szalay2015ict}. Originally, CAN refers to the ISO standard 11898-1 specifying the physical and data link layers of the CAN protocol stack. Second, another meaning is connected to FMS-CAN (Fleet Management System CAN), originally initiated by major truck manufacturers, defined in the SAE standard family J1939; FMS-CAN gives a full-stack specification including recommendations on higher protocol layers. Third, CAN refers to the multitude of proprietary CAN protocols which are make and model specific. This results in different message IDs, signal transformation parameters and encoding. These protocols are usually based on the standardized lower layers, but their higher layers are kept confidential by OEMs. The overwhelming majority of cars use one or more proprietary CAN protocols. Generally, sensor signals in CAN variants have a sampling frequency in the order of 10 ms.

On the other hand, using the standard on-board diagnostics (OBD, OBD-II) is a popular way of getting data out of the car. Originally developed for maintenance and technical inspection purposes and included in every new car since 1996, OBD is also used for telematics applications. Adding to the confusion regarding CAN, OBD has five minor variations including one which is based on the CAN physical layer. Sensor signals carried by OBD have a sampling frequency in the order of 1 second. In certain vehicle makes and models, one or more CANs are also connected to the OBD2-II diagnostic port. In such cars, also utilizing OBD over the CAN physical layer, it is possible to extract fine-grained CAN data via an OBD-II logger device.

Table \ref{tab:can_msg} shows a simplified picture of a CAN message with a 11-bit identifier, which is the usual format for everyday cars; trucks and buses usually use the extended 29-bit version. This example shows an already stripped message, i.e., we do not discuss end of frame or check bits.

\begin{table*}[tb]
    \caption{Example of CAN messages}
    \label{tab:can_msg} \centering
\begin{tabular}{|c|c|c|c|c|c|c|c|c|c|c|c|}
  \hline
  Timestamp & CAN-ID & Request & Length & Data \\
  \hline
  1481492683.285052 & 0x0208 & 000 & 0x8 & 0x00 0x00 0x32 0x00 0x0e 0x32 0xfe 0x3c \\
  1497323915.123844 & 0x018e  &  000  &  0x8 &  0x03 0x03 0x00 0x00 0x00 0x00 0x07 0x3f \\
  1497323915.112910 &  0x00f1 & 000 & 0x6 & 0x28 0x00 0x00 0x40 0x00 0x00 \\
  \hline
\end{tabular}
\end{table*}

\noindent \textbf{Components of a CAN bus message.}
\begin{itemize}
    \item Timestamp: Unix timestamp of the message
    \item CAN-ID: contains the message identifier - lower values have higher priority (e.g. wheel angle, speed, ...)
    \item Remote Transmission Request: allows ECUs to request messages from other ECUs
    \item Length: length of the Data field in bytes (0 to 8 bytes)
    \item Data: contains the actual data values in hexadecimal format. The Data field needs to be broken to sensor signals, transformed and/or converted to a human-readable format in order to enable further analysis.
\end{itemize}

Throughout this paper, we focus on the three practically relevant fields: CAN-ID, Length and Data.


\section{\uppercase{Data collection}}
\label{sec:data}

As CAN data logs are not widely available, we conducted a measurement campaign. For data collection in particular we connected a logging device to the OBD-II port and logged all observed messages from various ECUs. Such a device acts as a node on the CAN bus and is able to read and store all broadcasted messages. Our team developed both the logging device (based on a Raspberry PI 3) and the logging software (in C). Note that it is common that the OBD2 connector is found under the steering wheel. Also note that not all car makes and models connect the CAN serving the drive-train ECUs (or any CAN) to the OBD-II port (e.g., Volkswagen, BMW, etc.); in this case we could not log any meaningful data.

We have gathered meaningful data from 8 different cars and a total number of 33 drivers. We did not put any restriction on the demographics of the different drivers or the route taken. In each case we asked the driver to drive for a period of 30-60 minutes, while our device logged data from every route the drivers took. Drivers were free to choose their way, but still conforming to three practical requirements: (1) record at least 2 hours of driving in total, (2) do not record data when driving up and down on hills or mountains, (3) do not record data in extremely heavy traffic (short runs and idling). Free driving was recorded for all 33 drivers with an Opel Astra 2018: 13 people were between the age of 20-30, 12 between 30-40, and 8 above 40; there were 5 women and 28 men; 11 with less experience (less than 7000 km per year on average or novice driver), 9 with average experience (8-14000 km per year), and 13 with above average experience (more than 14000 km per year). 

We gathered data from the following cars: Citroen C4 2005 (22 message IDs), Toyota Corolla 2008 (36 IDs), Toyota Aygo 2014 (48 IDs), Renault Megane 2007 (20 IDs), Opel Astra 2018 (72 IDs), Opel Astra 2006 (18 IDs), Nissan X-trail 2008 (automatic, 34 IDs) and Nissan Qashqai 2015 (60 IDs). We would like to emphasize that the two Opel Astras use completely different prorpietary CAN versions (even the only 2 common IDs correspond to completely different Data). We also recorded the GPS coordinates via an Android smartphone during at least one logged drive per car. Most routes were driven inside or close to Budapest; approximately 15-20\% was recorded on a motorway.


\section{\uppercase{CAN data analysis}}
\label{sec:analysis}

All recorded messages contained 4 to 8 bytes of data; this made it likely that multiple (potentially unrelated) pieces of information can be sent under the same ID. We first assumed that signals are positioned over whole bytes; this turned out to be wrong. Our investigation revealed that besides signal values a message can also contain constants, multi-value fields and counters. Some values appear only on-demand, such as windscreen or window signals. All data apart from sensor signals are considered noise and, therefore, need to be removed.

Meaningful CAN IDs vary significantly across vehicle makes and models, therefore we expected that the only signals found in all cars with high probability are the basic ones: such as velocity, brake, clutch and accelerator pedal positions, RPM (round per minute) and steering wheel angle. Next, we devise a method that yields a deeper understanding of the Data field in CAN messages and a possibility for sensor signal extraction. Note that from this point we will use the term ID as a reference to both a given type of message and its data stream (time series).

\subsection{Bit decomposition heuristics}
\label{sec:bit}
Extracting the signals from a CAN message is not a trivial challenge. While monitoring the data stream while driving and finding the exact bits that change in reaction to one's actions is possible, it is highly time consuming, does not scale with hundreds of different existing CAN protocol versions and bound to miss out on potential sensor signals. (We only took this approach with a single car model to generate training data and a validation framework for our machine learning solution.) Our objective here is to present  our observations on message types and distributions that leads to a smarter message decomposition method.

First, we examined the message streams literally bit-by-bit. We presumed that inside a given ID with potentially multiple sensor readings there was a difference in their bit value distribution, hence they could be systematically located and partitioned according to some rule. E.g., let us assume that there are two signals sent next to each other under the same ID (i.e., there are no zero bits or other separators between the two. Given that signals are encoded in a big endian (little endian) format, both of their MSBs (LSBs) are rarely $1$s. Therefore, there should be a drop in bit probability (i.e., the probability for a given bit to be $1$) between the last bit of the first signal and the first bit of the second signal. In order to visualize these drops we represent IDs by their bit distribution: we sum the number of messages for each ID and how many times a given bit was one and divide these two measures:
\begin{center}
    $v_{i} =  \frac{\sum_{j=1}^{|v|} \mathbbm{1}_{\{ v_j=1\}}}{|v|} $ 
\end{center}

where $v$ denotes the binary vector of a given ID, that is the representation of a CAN message's payload in binary format, and where $v_{i}$ denotes the probability of a bit being 1 at the i$^{th}$ position.

\begin{figure}[h]
\centering
\subfigure[Driver A]{\label{fig:bit_dif1}\includegraphics[width=\linewidth]{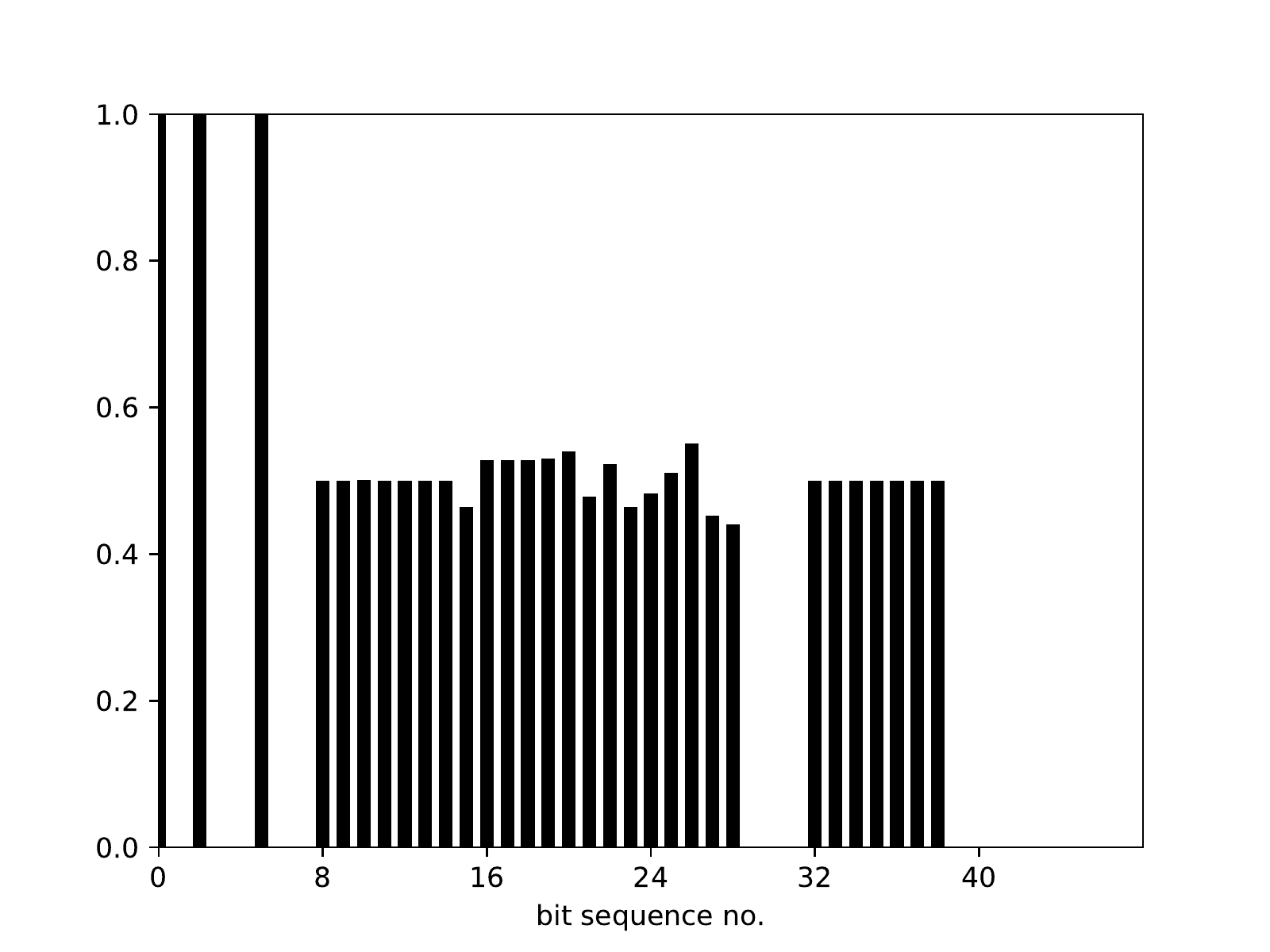}}
\subfigure[Driver B]{\label{fig:bit_dif2}\includegraphics[width=\linewidth]{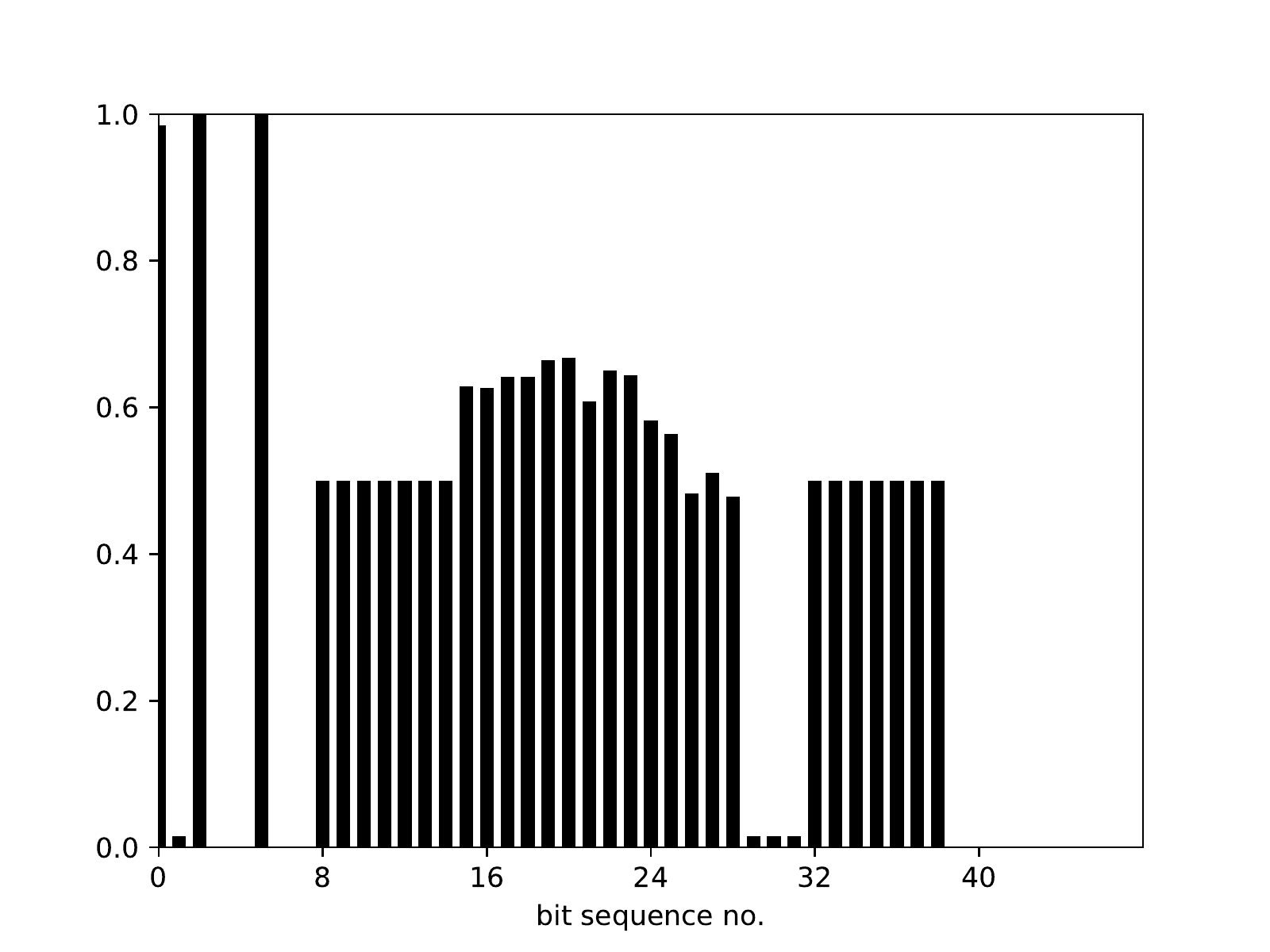}}
\caption{Bit distribution of the same ID from different drivers}
\end{figure}

When we examined the distribution of the bits in an ID we found that that in some cases it is straightforward to extract a signal: between two signal candidates there were separator bits with $v_{i} = 0$ or $v_{i} = 1$. Other cases were more complex: given Figure \ref{fig:bit_dif1} it is hard to determine signal borders. However, combined with the bit distribution from the same ID and car model but another drive, the signals became clearly distinguishable. 

\subsection{Pre-processing}
\label{sec:preprocessing}
 
\begin{figure}[h]
\centering
    \includegraphics[width=\linewidth]{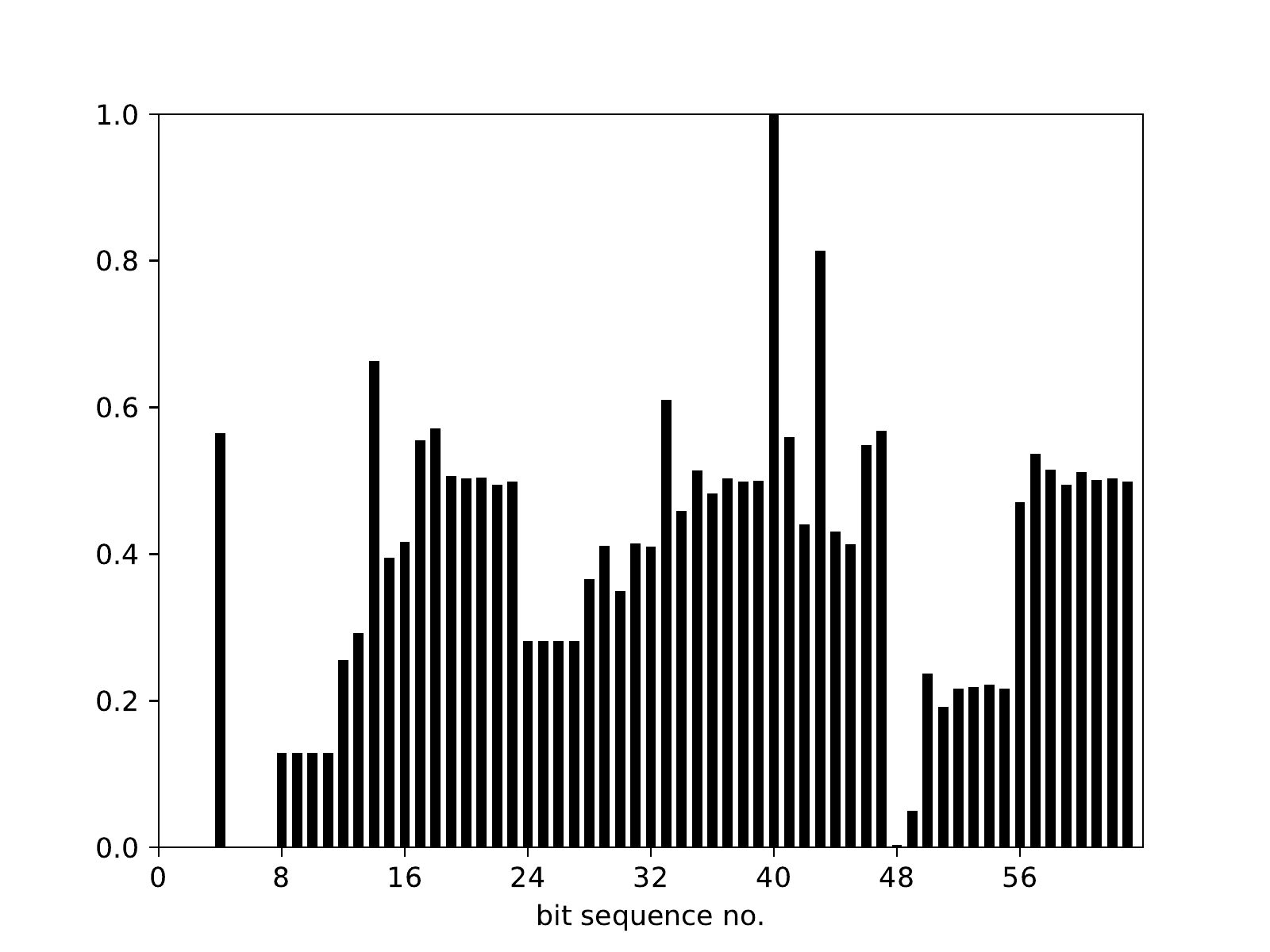}
    \caption{Consecutive blocks in a CAN message that cannot be unambiguously divided}
    \label{fig:bit_blocks}
\end{figure}

After examining bit distributions we realized that $\approx 90\%$ of candidate signal blocks are placed on one or two bytes. In other cases signal borders were not unambiguous, see Figure \ref{fig:bit_blocks}. Our first heuristic suggests a start of a new signal because of the drop at the 23rd and the 24th bits, although it is clearly a counter or a constant on 3 bits, but we can not determine where exactly a new signal starts (is it the 28th bit or the 32nd?). Moreover, the 41th bit is constant $1$ bit which might signify some kind of a separator, yet we cannot be certain. After a long evaluation we decided to divide the data part of the messages to bytes and pairs of bytes; as a result for one ID we could define 4 to 8 sensor candidates.

\noindent \textbf{Filtering.}
Examining the byte time series resulting from the above approach, we spotted that many series were constant, had very few values, were cyclic (counters) or changed very rarely. As we intended to use machine learning to find the exact signals, not filtering these samples could have caused significant performance loss and a bloated and skewed training dataset with a lot of similar negative samples resulting in a decreased variability of training data potentially to a degree of corrupting the model. Therefore, we evaluated the variation for each sample and excluded those that had a very low variation ("low variation" was also a free variable optimized during evaluation).

\noindent \textbf{Normalization. }
We scaled all candidate time series to the interval of $[0,1]$: we extracted the maximum value for the whole candidate series, then we divided all values by the maximum. Scaling the data solves the problem of transformed (shifted) values, i.e., the same signal can take different values during drives, that can be a result of some transformation on the data in one vehicle or simply the fact that one car was driven in a lower range of velocity in contrast to the other (i.e., one log comes from a drive that did not exceed 50 km/h while the other rarely drove slower than 100 km/h).

\noindent \textbf{Sliding windows. }
We divided logs into overlapping sliding windows from which we extracted our features for machine learning. The sliding window length (elapsed time) and percentage of overlap with previous and successive windows were free variables which we set to default values and subsequently optimized during run time.

\section{Classification}
\label{sec:classification}
We use machine learning for two purposes; first, to extract signals from CAN messages, and second, to perform driver re-identification using the extracted signals. Therefore, we build two different types of classifiers. In order to extract signals, we train a classifier per signal on the statistical features of the signal in a base car (e.g., Opel Astra 2018) where we exactly know where the signal resides, that is, the message ID and byte number of the CAN message which contains the signal. Then, we use the trained models to identify the same signals in another car (e.g., Toyota) where the locations of the signals (i.e., message ID and byte number) are unknown. The intuition is that the physical phenomenon that a signal represents has identical statistical features irrespective of the car, and hence can be used to identify the same signal in all cars using the same classifier.

For driver re-identification, similarly to previous works \citep{drivers_classification,GMM_driving}, we use a separate classifier that is trained on the already extracted signals of the car. This classifier learns the distinguishing features of different drivers (and not that of signals like the first classifier) using the signals produced during their drives.  
 
For both signal extraction and driver re-identification, the features computed from each sliding window constitute a single training sample (i.e., a sample vector) used as the input of our machine learning classifiers. Below we describe the classifiers, the division of training and testing samples, and the method used for multi-class classification.

\subsection{Multi-class Classification} 
\label{sec:multicl}
We implemented multiclass classification using binary classification in a one-vs-rest way (aka, one-vs-all (OvA), one-against-all (OAA)). The strategy involves training a single classifier per class, with the samples of that class as positive samples and all other samples as negatives. For signal extraction, a class represents a pair of message ID and byte number, whereas for driver re-identification, it represents a driver's identity.
A random forest model was trained per class with balanced training data (i.e., containing the same number of positive and negative samples), and its output was binary indicating whether the input sample belongs to the class or not. 
For signal extraction, as each training/testing sample is a small portion of the time-series  (i.e., window) representing a signal, we apply the trained model on all portions of a signal and obtain multiple decisions per signal. Then, the "votes" are aggregated and the  candidate signal with the most number of "votes" is selected. 
  
We would like to stress that random forests are indeed capable of general multiclass classification without its transformation to binary. We have also tried this general multiclassification approach, however, its results were inferior to the OvA's results. Moreover successful driver re-identification can already be carried out using a single or only a few signals \cite{fingerprint}. In this paper, we use the velocity, the brake pedal, the accelerator pedal, the clutch pedal and the RPM signals to extract for driver re-identification. 

\begin{table}
\caption{Average feature importances}
\label{tab:feature_importances}
\centering
\begin{tabular}{|c|c|c|}
  \hline
    Feature & Importance\\
    \hline
    count below mean & 0.2113 \\
    count above mean & 0.1482\\
    cid\_ce & 0.1290\\
    mean abs change & 0.1048\\
    maximum & 0.0716\\
    longest strike below mean & 0.0708\\
  \hline
\end{tabular}
   
\end{table}
\subsection{Feature extraction}
\label{sec:feature}
Our classifiers use statistical features of the samples; for each sliding window we extracted 20 different statistics that are widely considered as most descriptive regarding time series characteristics (see the best features in Table \ref{tab:feature_importances}. 
We finally used 15 features based on their importances calculated from our random forest models. These features are the following:
\begin{enumerate}
    \item $count\_above\_mean(x)$: Returns the number of values in $x$ that are higher than the mean of $x$.
    \item $count\_below\_mean(x)$: Returns the number of values in $x$ that are lower than the mean of $x$.
    \item $longest\_strike\_above\_mean(x)$: Returns the length of the longest consecutive subsequence in $x$ that is bigger than the mean of $x$.
    \item $longest\_strike\_below\_mean(x)$: Returns the length of the longest consecutive subsequence in $x$ that is smaller than the mean of $x$.
    \item $binned\_entropy(x, max\_bins)$: First bins the values of $x$ into $max\_bins$ equidistant bins. The $max\_bin$ parameter was generally set to 10. Then calculates the value of:
    \begin{center}
        $- \sum_{k=0}^{min(max\_bins,len(x))} p_k \cdot log(p_k) \cdot \textbf{1}_{\{p_k > 0\}}$
    \end{center}
    where $p_k$ is the percentage of samples in bin $k$.
    \item $mean\_abs\_change(x)$: Returns the mean over the absolute differences between subsequent time series values which is:
    \begin{center}$\frac{1}{n} \sum_{i=1,\ldots, n-1} | x_{i+1} - x_{i}|$    \end{center}
    \item $mean\_change(x)$: Returns the mean over the differences between subsequent time series values which is:
    \begin{center} $\frac{1}{n} \sum_{i=1,\ldots, n-1} (x_{i+1} - x_{i})$\end{center}
    \item OTHER: minimum, maximum, mean, median, standard variation, variance, kurtosis, skewness
\end{enumerate}
This way we created an input vector of features for each sample (one sample corresponds to one window). No smoothing, outlier elimination or function approximation are performed on the samples before feature extraction. For calculating the above statistics, we used the $tsfresh$ python package\footnote{\url{https://tsfresh.readthedocs.io/en/latest/}} .

\subsection{Training and model optimization}
\label{sec:training}

For training our classifier we need to have a ground truth of sensor signals from a single car. These certified signals then can be compared to the candidate signals from other cars to find the best match. We chose the Opel Astra 2018 as our reference, as we had the most drives logged from this car.


\noindent \textbf{Velocity versus GPS. }
We recorded GPS coordinates for all drives with the Opel Astra 2018. Setting the Android GPS Logger app to the highest accuracy (complemented by cell tower information achieving an accuracy of 3 meters) and saving the coordinates every second, we ended up with a time series of locations. Using the timestamps, GPS time series also determines the mean velocity between neighboring locations, producing a velocity time series. Intuitively, the GPS based velocity is very close to the one recorded from the CAN bus. 

In order to test this hypothesis we applied the Dynamic Time Warp algorithm (DTW)~\cite{DTW1}. The DTW algorithm is part of time series classification algorithms \cite{great_ts}, their important characteristic being that there may be discriminatory features dependent on the ordering of the time series values \cite{pattern_extr}. A distance measurement between time series is needed to determine similarity between time series and for classification. Euclidean distance is an efficient distance measurement that can be used. The Euclidean distance between two time series is simply the sum of the squared distances from each $n^{th}$ point in one time series to the $n^{th}$ point in the other. The main disadvantage of using Euclidean distance for time series data is that its results are very un-intuitive. If two time series are identical, but one is shifted slightly along the time axis, then Euclidean distance may consider them to be very different from each other. DTW was introduced to overcome this limitation and give intuitive distance measurements between time series by ignoring both global and local shifts in the time dimension. DTW finds the optimal alignment between two time series if one time series may be warped non-linearly by stretching or shrinking it along its time axis.

Before running DTW we excluded the outliers from the GPS-based velocity series. These points are the result of GPS measurement error and materialize in extreme differences between two neighboring velocity values (we used 30 km/h as a limit). We then ran DTW with the GPS-based velocity values against all other sensor candidates of the CAN log. As the result of the DTW algorithm is a distance between two series, the smallest distance yields the best match: in every case it was indeed the same ID by a wide margin (see Table \ref{tab:dtw_velo_opel}). We used manual physical tryouts to corroborate that this ID indeed corresponds to velocity.

\begin{table}[tb]
\centering
\caption{DTW velocity search top results}
\label{tab:dtw_velo_opel} 
\begin{tabular}{|c|c|c|}
  \hline
    CAN ID & Byte & Distance\\
    \hline
    0410	& 1-2 & 7499\\
    0410	& 2-3 & 15972\\
    0295 & 1-2 & 20609\\
    0510 & 2 & 20981\\
    0510 & 3 & 21585\\
  \hline
\end{tabular}
\end{table}

\begin{figure*}[tb]
    \centering
    \subfigure[Brake pedal (black) vs. accelerator pedal (red) position]{\label{fig:brake_vs_acc}\includegraphics[width=0.9\linewidth]{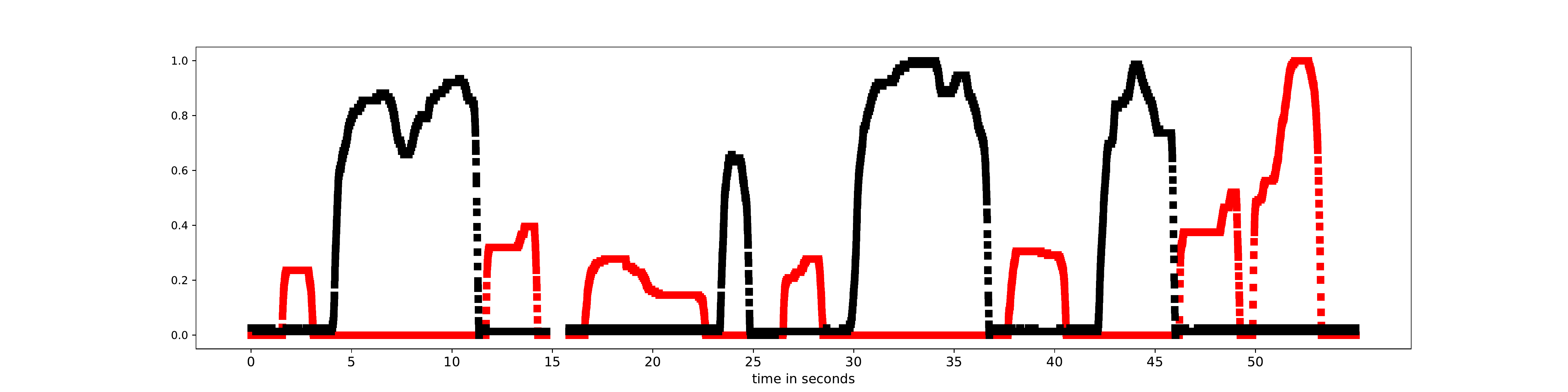}}
    \subfigure[A false match]{\label{fig:brake_false}\includegraphics[width=0.9\linewidth]{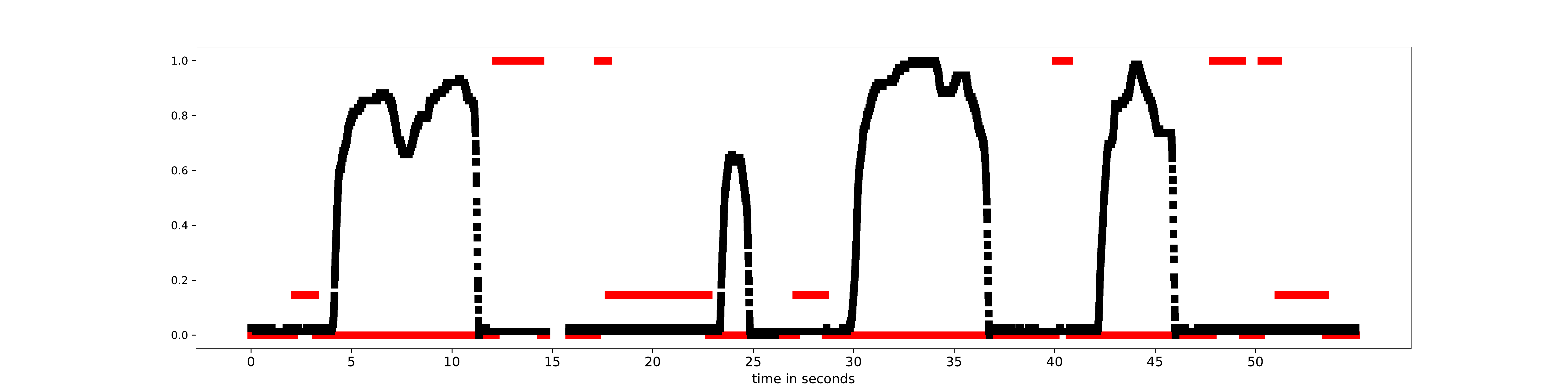}}
    \caption{Searching for the brake and accelerator pedal position signals}
    \label{fig:brakevs}
\end{figure*}

\noindent \textbf{Brake vs. accelerator: pedal position. }
Extracting the brake and the accelerator pedal positions required a different approach. In a normal vehicle the accelerator and the brake pedal are not pressed at the same time because it contradicts a driver's normal behaviour (excluding race car drivers). Consequently, to extract the accelerator and the brake pedal positions one only have to search for a pair of signals that are almost exclusive to each other. For this end,  we compared all pairs of ID byte subseries from multiple drives and listed the candidates that fit the description. Figure \ref{fig:brake_vs_acc} shows the correct result and Figure \ref{fig:brake_false} shows false candidate. False results were easy to exclude because of their characteristics;  in this example it is trivial that a piece-wise constant signal cannot possibly signify a pedal position. Finally, we used manual physical tryouts to corroborate that these IDs indeed correspond to the brake and accelerator pedal positions, respectively. Note that older vehicles can have a binary brake (and clutch) signal, as there is no corresponding sensor signal in them. 

\begin{figure*}[tb]
\centering
  \includegraphics[width=0.9\linewidth]{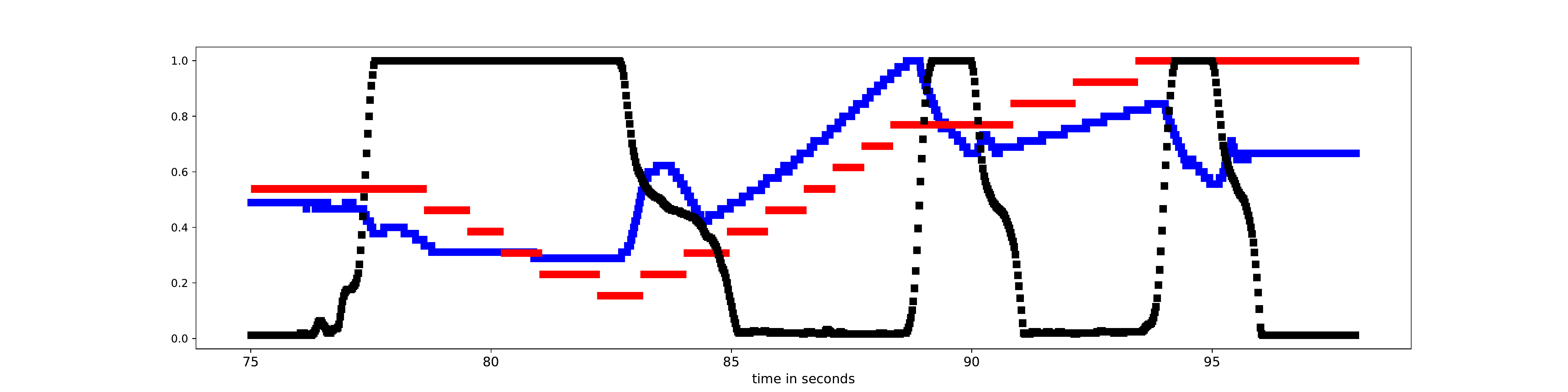}
  \caption{The clutch pedal position (black) vs. RPM (blue) vs. velocity (red) of one of our test vehicles.}
  \label{fig:clutch}
\end{figure*}

\noindent \textbf{Clutch vs. RPM vs. velocity. }
The clutch pedal position also has very typical characteristics especially when compared with the velocity and RPM values. Once we start to accelerate from $0$km/h usually we change the gears quickly, thus the changes in the rpm and clutch pedal position are easy to detect. Upon gear change the RPM drops, then rises as we accelerate, then drops and rises again until we reach the desired gear and velocity. During the same time we push the clutch pedal every time just before the gear is changed. Moreover, we tend to use the clutch pedal in a very typical way, when the driver releases the clutch there is a slight slip around the middle position of the pedal indicating that the shafts start to connect. (Note that the length of this slip is characteristic for car models, condition (e.g., bad clutch) and driver experience.) Applying this common knowledge we searched for a pair of signals with one of them having a sharp spike (RPM) and the other a small platform (slipping clutch) around the same time. We narrowed our search to cases when the vehicle accelerated from zero to at most 50 km/h. In Figure \ref{fig:clutch} we can see these signal characteristics compared to each other. We managed to find the clutch pedal position and RPM signals based on the above. As before, we validated our findings with manual physical tryouts.\\
\noindent \textbf{Optimization. }
After extracting the ground truth signals, we calculated the feature vectors and trained a random forest classifier for each extracted signal: velocity, brake pedal position, accelerator pedal position, clutch pedal position and engine RPM. For parameter optimization and testing we tested our model on logs from the same car, but driven by another driver on another route. 
\subsection{Results}
\label{sec:results}
Next we describe the performance of our classifiers used to extract signals from CAN logs (in Section \ref{sec:signal_extraction}) and to re-identify drivers using the extracted signals (in Section \ref{sec:reid}).

\subsubsection{Signal extraction}
\label{sec:signal_extraction}
Our random forest classifiers used for signal extraction are trained on the CAN logs of a base car (here it is an Opel Astra'18) where the locations of a target signal is known. The classifiers take statistical data vectors as inputs with the 15 statistical features (see in Section \ref{sec:feature}) extracted from each sample (window). In particular, we train a random forest classifier to distinguish a target signal from all other signals, where the positive training samples are composed of the windows of the time series corresponding to the target signal, whereas negative samples are taken from other signals' time series. Hence, we obtain a classifier per target signal.
Recall that signal locations are computed using the techniques described in Section \ref{sec:training}.
We apply each trained classifier on all the samples (windows) of all time series in another (target) car where we want to locate the corresponding target signals. For every classifier, we obtain a classification for each window of each time series in the target car. The time series which receives the largest number of votes (i.e., has the most windows classified as positive) will be the matched signal, i.e. the signal which is the most similar to the target signal.

Best results were obtained using the following parameter settings: the length of windows is set to 2.5 seconds, which is sufficiently large to capture different driver reactions (one can accelerate from 0 to even 30 km/h or can hit the brakes and stop the vehicle). The sampled logs are at least 30 minutes long, the overlap parameter is set to 25\%.  The pruning parameter is set to 7, i.e. a sample was excluded when its variation is less than 7. 

Each trained random forest classifier is tested against samples from logs of all other cars except the base car, and the logs were pre-processed as described in Section \ref{sec:preprocessing}. The matching performed by a classifier is validated by manually extracting the ground truth sensor signal from the target car as described in Section \ref{sec:training}. 
In order to measure the accuracy of our classifier, we report the rank of the true signal; each candidate signal is ranked according to the number of votes (i.e., positive classifications) they receive, i.e. the signal having the highest vote ranked first.

Table \ref{tab:match} shows the results of signal extraction using only 30 minutes of data for training and also 30 minutes for matching (testing), where training was performed on CAN logs obtained from our base car (i.e., Opel Astra'18). Three signals (RPM, velocity and accelerator pedal position) are all ranked in the first place (i.e. received the highest number of votes in the classifier),that is, \emph{our approach successfully identified all three signals in all the target cars}. Note that we did not exract the clutch and the brake pedal position signals as during the validation we realized that these signals do not even exist in most of our cars in the database.

We also report the precision (TP/TP+FP) and recall (TP/TP+FN) of the classifier (where TP=true positive, FP=false positive and FN=false negative) which represent how many positive classifications are correct over all samples of all time-series in the CAN log (precision), and how many samples of the true matching signal are correctly recognized by the classifier (recall). We also compute and report the gap which is the difference between the number of votes (i.e., positive classifications) of the highest ranked true signal and that of the highest ranked false signal divided by the total number of votes. For example if the highest ranked true signal received 50\% of the votes and the highest ranked false signal received 20\%, then the gap equals $0.30$. Note that in most cars most sensors appear under several IDs in the log, this causes that more than one candidate with very high votes are all true positives, thus the top high ranks can all be true positives and the highest ranked false signal drops to the fourth, fifth or even lower places.

\begin{table}
\small
\caption{Top results against RPM, velocity and acceleration}
\label{tab:match}
\centering
\begin{tabular}{|c|c|c|c|c|}
  \hline
    \small
    Sensor & \textbf{Rank} & Precision & Recall & Gap\\
    \hline
    Citroen\_rpm & \textbf{1} & 0.353 & 0.952 & 0.180\\
    Citroen\_velo & \textbf{1} & 0.874 & 0.792 & 0.094\\
    Citroen\_acc & \textbf{1} & 0.740 & 0.444 &  0.431\\
    \hline
    Opel\_A06\_rpm & \textbf{1} & 0.155  & 0.604 & 0.035\\
    Opel\_A06\_velo & \textbf{1} & 0.158 & 0.969 & 0.024\\
    Opel\_A06\_acc & \textbf{1} & 0.207 & 0.934 & 0.026\\
    \hline 
    Toyota\_A\_rpm & \textbf{1} & 0.229 & 0.717 & 0.238\\
    Toyota\_A\_velo & \textbf{1} & 0.230 & 0.214 & 0.337 \\
    Toyota\_A\_acc & \textbf{1} & 0.394 & 0.566 & 0.296\\
    \hline
    Toyota\_C\_rpm & \textbf{1} & 0.224 & 0.399 & 0.049 \\
    Toyota\_C\_velo & \textbf{1} & 0.676 & 0.139 & 0.029 \\
    Toyota\_C\_acc & \textbf{1} & 0.602 & 0.522 & 0.134 \\
    \hline
    Renault\_rpm & \textbf{1} & 0.439 & 0.991 & 0.148 \\
    Renault\_velo & \textbf{1} & 0.890 & 0.522 & 0.465 \\
    Renault\_acc & \textbf{1} & 0.491 & 0.702 & 0.210\\
    \hline
    Nissan\_X\_rpm & \textbf{1} & 0.248 & 0.805 & 0.140\\
    Nissan\_X\_velo & \textbf{1} & 0.970 & 0.870 & 0.466\\
    Nissan\_X\_acc & \textbf{1} & 0.484 & 0.728 & 0.527\\
  \hline
    Nissan\_Q\_rpm & \textbf{1} & 0.211 & 0.680 & 0.034 \\
    Nissan\_Q\_velo & \textbf{1} & 0.462 & 0.522 & 0.110 \\
    Nissan\_Q\_acc & \textbf{1} & 0.512 & 0.774 & 0.044 \\
    \hline
\end{tabular}
\end{table}

\subsubsection{Driver re-identification}
\label{sec:reid}
Next we use the extracted signals in a driver re-identification scenario. We use the same preprocessing as in Section \ref{sec:preprocessing} and the same parameter settings as in Section \ref{sec:training}, except that we do not use all 15 features, only 11 of them are chosen based on their importances:  \textit{count above mean, count below mean,
longest strike above mean, longest strike below mean, maximum, mean, mean abs change, median, minimum, standard  deviation, variance}. We used four extracted signals: accelerator and brake pedal positions, velocity and RPM. The feature vector of a driver consists of  44 features altogether. All drivers used the same car, which was Opel Astra'18, to produce CAN logs. The samples were divided into a training and testing set, where the training and testing data made 90\% and 10\% of all samples, respectively. We used 10-fold cross-validation to evaluate our approach. We selected 5 drivers uniformly at random, and built a binary classifier for each pair of drivers. Our classifier achieved 77\% precision on average (each model was evaluated 10 times). The worst result was just under 70\% and the best result was 87\%.

\section{Conclusion and Future Work}
\label{sec:conclusion}
We described a technique to extract signals from vehicles' CAN logs. Our approach relies on 
using unique statistical features of signals which remain mostly unchanged even between different types of cars, and hence can be used to locate the signals in the CAN log.  
 We demonstrated that the extracted signals can be used to effectively  identify drivers in a dataset of 33 drivers. 
 Although our results need to be evaluated on a larger and more diverse dataset, our findings show that driver re-identification can be performed without the nuisance of signal extraction or agreements with a manufacturer. This means that not revealing the exact signal location in CAN logs is not sufficient to provide any privacy guarantee in practice. 
Car companies should devise more principled (perhaps cryptographic) approaches to hide signals, and/or to anonymize their CAN logs so that drivers cannot be re-identified. 
\section*{\uppercase{Acknowledgements}}
This work has been partially funded by the European Social Fund via the project EFOP-3.6.2-16-2017-00002, by the European Commission via the H2020-ECSEL-2017 project SECREDAS (Grant Agreement no. 783119) and the Higher Education Excellence Program of the Ministry of Human Capacities in the frame of Artificial Intelligence research area of Budapest University of Technology and Economics (BME FIKP-MI/FM). Gergely Acs has been supported by the Premium Post Doctorate Research Grant of the Hungarian Academy of Sciences (MTA). Gergely Bicz{\'o}k has been supported by the J{\'a}nos Bolyai Research Scholarship of the Hungarian Academy of Sciences.


\bibliographystyle{apalike}
{\small
\bibliography{example}}

\end{document}